\documentclass[aps,pra,twocolumn,groupedaddress]{revtex4-1}
\usepackage[pdftex]{graphicx}
\usepackage{float}
\usepackage{amsmath}
\usepackage{amssymb}
\usepackage{url}
\usepackage{epsfig}
\usepackage{epstopdf}

\begin{document}

\title{Survival of the fittest in the coherent evolution of quantum ensembles}

\author{G. Liu, O. Be'er, Y. Margalit, M. Givon, D. Groswasser, Y. Japha}
\email[Correspondence should be addressed to:\\]{japhay@bgu.ac.il}
\author{R. Folman}
\affiliation{Department of Physics, Ben-Gurion University of the Negev, Beer-Sheva 84105, Israel.}
\date{\today}

\begin{abstract}
We report two novel effects in an inhomogeneous ensemble of two-level systems driven by an external field. First, we observe a rigidity of the oscillation frequency: the dominant Rabi oscillation frequency does not change with the frequency of the driving field, in contrast to the well-known law of Rabi frequency increase with growing detuning of the driving field.  Second, we observe a time-dependent frequency shift of the ensemble-averaged oscillation. We show that these effects follow from the inhomogeneity of the two-level splitting across the ensemble, allowing for a distribution of local oscillations in which those with high frequencies interfere destructively and decay faster than those with a low frequency, which are the only to survive in the output signal.
Hence, coherence emerges from long-lived oscillations in an inhomogeneous ensemble. We analyze the Fourier spectrum of the time-dependent oscillation signal and find a non-trivial spectral structure that is double peaked for certain parameters.
We show that the effects observed in alkali vapor are universal and expected in any system with a moderate inhomogeneity driven by an external field.
\end{abstract}

\maketitle

\section{Introduction}

Rabi oscillations are an elementary process which is studied in a wide range of two-level systems driven by near resonant electromagnetic fields in the optical \cite{Knight1980rabi}, micro-wave and radio-frequency regimes \cite{RevModPhys.76.1037}. This process is of immense fundamental and technological significance \cite{budker2007optical, andrew2013jaynes, PhysRevLett.93.130501, press2008complete,
 palacios2010experimental, dudin2012observation, vijay2012stabilizing, PhysRevLett.111.073001} and consequently it is important that all its aspects be well understood. According to theoretical models, such as the Rabi or Jaynes-Cummings formalisms \cite{andrew2013jaynes}, the response of the system is described by the generalized Rabi frequency rule, stating that the oscillation frequency grows while the amplitude drops with increasing detuning of the driving field from the natural resonance of the two-level system. However, uncertainties and inhomogeneities in both the system and the field increase the complexity of the situation, giving rise to a rich spectrum of features.

Here we characterize and explain unique aspects of an inhomogeneous ensemble of quantum two-level systems. Such ensembles are common in vapor, cold atom clouds, Bose-Einstein condensates, arrays of traps for cold ions and atoms, as well as related systems such as  nitrogen-vacancy centers in diamonds.
One noticeable effect of inhomogeneity in such systems is dephasing leading to the decay of coherence. This effect is observed in many kinds of systems performing driven or natural oscillations (such as in Ramsey interferometry) and due to different kinds of inhomogeneities: variation of the local field or Doppler shifts due to a thermal velocity distribution.
A particular example is a cloud of cold atoms performing Rabi oscillations in an inhomogeneous driving field, where the spatial and temporal population pattern decays due to thermal velocity distribution~\cite{Anat_PRA2013}.

 \begin{figure}
\includegraphics[width=\columnwidth]{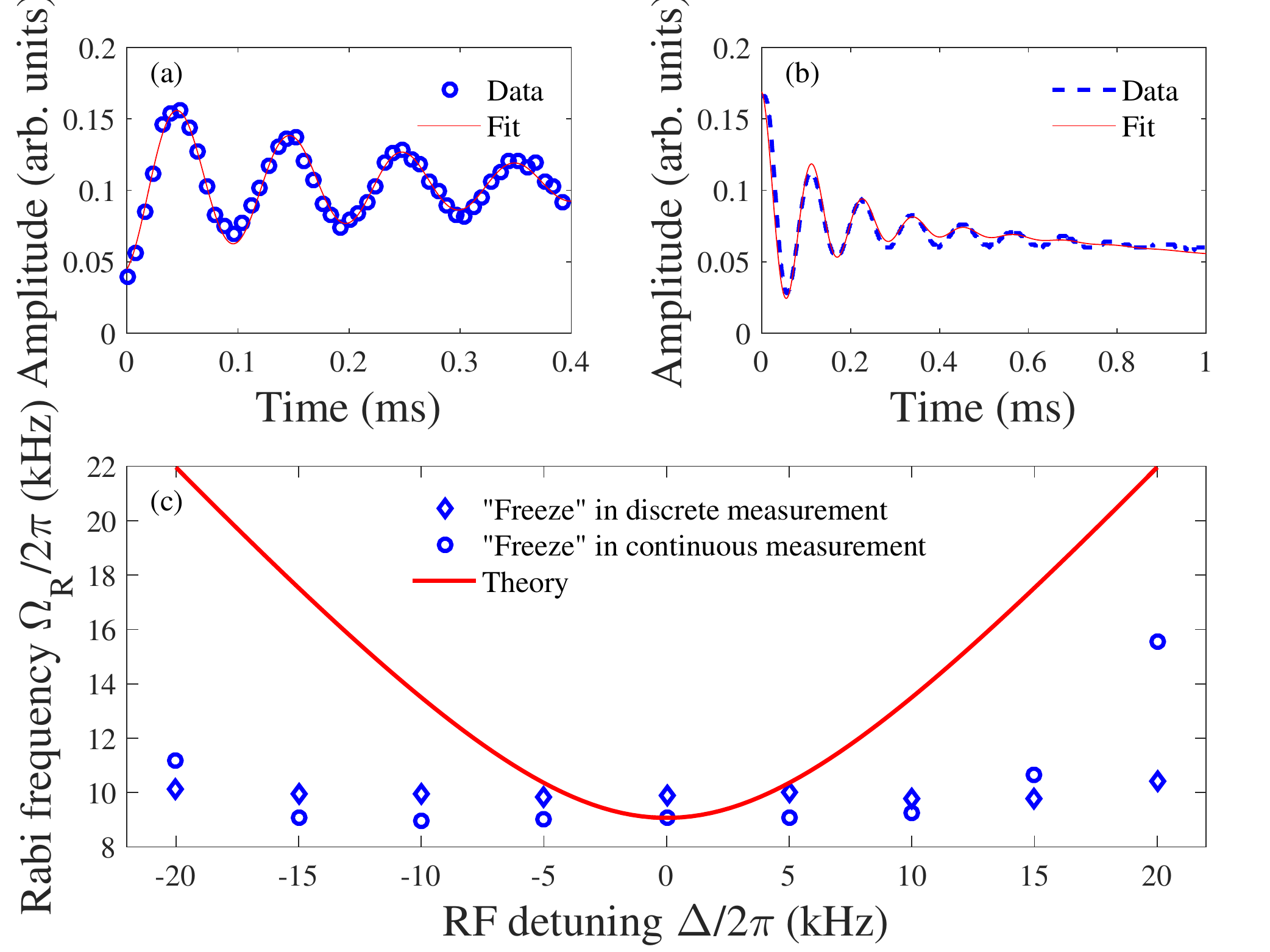}
\caption{(color online) Frequency rigidity of Rabi oscillations in a vapor cell (see section~\ref{sec:experiment}). (a) Discrete measurement: Typical Rabi oscillations driven by an RF amplitude corresponding to a bare Rabi frequency $\Omega_{0}/2\pi\sim$10\,kHz. (b) Continuous measurement: Typical Rabi oscillations with a bare Rabi frequency $\Omega_{0}/2\pi\sim$9\,kHz. Solid red lines in (a) and (b) are single-frequency fits. (c) Rabi frequency vs. RF detuning. Solid red line in (c) is the expected generalized Rabi frequency according to $\Omega_{\rm_R}=\sqrt{{\Omega_{0}}^{2}+\Delta^{2}}$, where $\Delta$ is the detuning. The rigidity is clearly visible.}
\label{fig:2methods}
\end{figure}

We report on novel experimental observations related to the frequency of Rabi oscillations in an ensemble of atoms in a vapor cell, a system which is still of significant interest for applications in quantum science and technology \cite{PhysRevLett.104.133601,krauter2013deterministic,PhysRevLett.110.123001}.
We have measured the frequency of Rabi oscillations between two Zeeman sub-levels induced by a radio-frequency (RF) magnetic field and found that the oscillation frequency barely changes with the driving field's frequency, as shown in Fig.~\ref{fig:2methods}. This rigidity of the oscillation frequency seems to contrast the well-known rule of generalized Rabi frequency, according to which the oscillation frequency should grow with increasing detuning of the driving field frequency from the resonant transition frequency of the atoms.
An analysis of this frequency rigidity, which we have observed for a wide range of parameters, shows that it originates from magnetic field inhomogeneity across the ensemble.
Although this inhomogeneity is relatively small (0.05\% or less) and difficult to improve, it generates spectral broadening comparable to the Rabi frequency and leads to a clearly manifested phenomenon.
To the best of our knowledge, such a phenomenon was not previously reported or predicted.

In addition to describing the observation in detail, we present a theoretical explanation and show that it is  general, suggesting that the same phenomenon should be observable in any system involving driven oscillations of a two-level system in the presence of inhomogeneous broadening, provided that the broadening is on the order of the oscillation frequency.
We use an abstract model to demonstrate the mechanism involved in the rigidity effect, which consists of destructive interference between oscillations far from resonance and survival of contributions to the overall signal from parts of the ensemble that oscillate near resonance.
Both the theoretical model and the experimental observations demonstrate some other features related to this mechanism.
 In particular, a shift from high to low frequency during the Rabi oscillation is predicted and clearly observed experimentally for certain parameters.
This effect is related to the structure of the frequency spectrum of the oscillation signal, which show a double-peak form for some parameters.

In what follows we describe the experimental results and provide a quantitative explanation based on a theoretical model of the experimental system. Next, in section~\ref{sec:theory}, we present a theoretical insight based on a simplified model that does not rely on specific properties of the experimental setup and emphasizes the universality of the effects.

\section{Experiment}
\label{sec:experiment}

\subsection{Experimental setup and procedure}
\label{sec:setup}

\begin{figure}
\centering
\includegraphics[width=\columnwidth,trim={3cm, 2.5cm, 2.5cm, 1cm},clip]{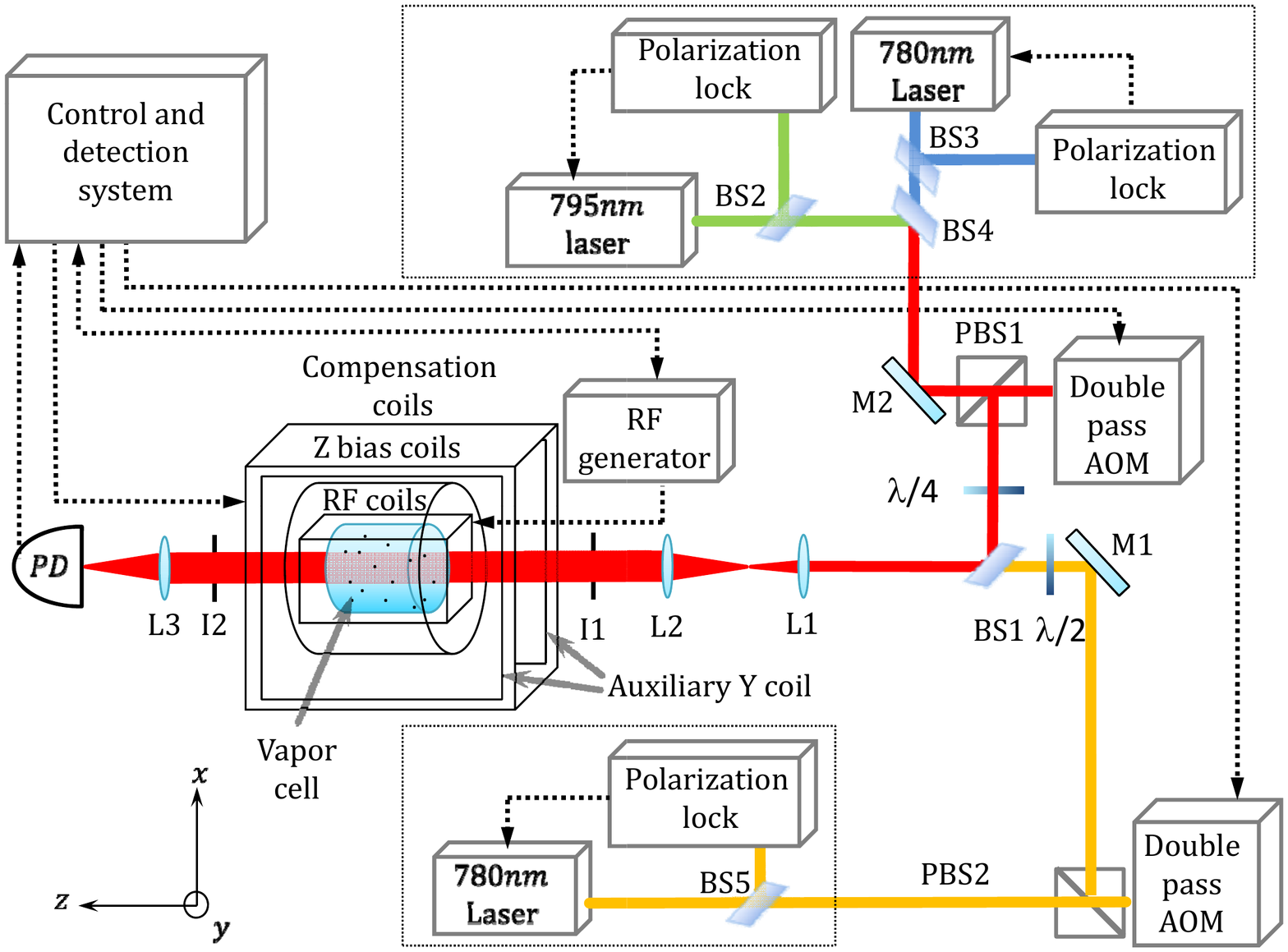}
s\caption{ (color online) Experimental setup. BS1, BS2, BS3, BS4 and BS5 are beam splitters; PBS1 and PBS2 are polarizing beam splitters; L1, L2 and L3 are lenses; I1 and I2 are irises; M1 is a mirror and PD is a photo-diode. Dashed lines indicate data and control lines. The double-pass AOM can turn the laser beams on and off within 1 $\mu$s. The lenses L1 and L2 and the iris I1 control the diameter and the intensity profile of the beam.}
\label{fig:setup}
\end{figure}

The experimental setup was presented previously  in \cite{PhysRevLett.111.053004}. It is depicted here in Fig.\,\ref{fig:setup}.
We induce Rabi population oscillations  in a $^{87}$Rb vapor contained in a cylindrical ($\oslash 25 \times 38\,$mm) cell with neon buffer gas at room temperature and 75\,Torr.

The two-level system used for the Rabi oscillation is a system of Zeeman sub-levels of the hyperfine level $F=2$ that were energetically split in a static magnetic field, produced by several sets of Helmholtz coils.
We exploit the nonlinear Zeeman effect to isolate two of the Zeeman sub-levels from the others, such that transitions are induced only between the sub-levels $|F,m_F\rangle=|2,2\rangle\equiv |2\rangle$ and $|2,1\rangle\equiv |1\rangle$. This is facilitated by a magnetic field of about 26\,G, strong enough to shift transitions other than the one under investigation by about 100\,kHz.
Three mutually perpendicular sets of square Helmholtz compensation coils cancel Earth's magnetic field.
The $z$ bias coils produce the 26\,G DC magnetic field parallel to the axis of the vapor cell, while the auxiliary $y$ coils produce a 1\,G magnetic field in the $y$ direction.

RF coils produce an oscillating magnetic field with amplitude $B_{\rm RF}$ perpendicular to the vapor cell axis (quantization axis) for driving transitions between the two Zeeman sub-levels. The driving rate (bare Rabi frequency) is $\Omega_0=g_F \mu_B\langle 1|\hat{F}_{\perp}|2\rangle B_{\rm RF}/\hbar$, where $\mu_B$ is the Bohr magneton, $g_F\approx 1/2$ is the Land\'e factor for the hyperfine level $F=2$ and $\langle 1|\hat{F}_{\perp}|2\rangle=1$ is the matrix element of the perpendicular component of the total angular momentum ${\bf F}={\bf L}+{\bf S}+{\bf I}$ between the two Zeeman sub-levels.
We limit $\Omega_0$ to a maximum of $25$\,kHz, thereby keeping the two-photon transition rate to $|2,0\rangle$ negligible.

Pumping the atoms into the desired levels is done by two circularly polarized lasers tuned to the F=1 $\leftrightarrow$ F'=2 transition in the D$_2$ line (780\,nm laser on the top of Fig.\,\ref{fig:setup}) and to the F=2 $\leftrightarrow$ F'=2 transition in the D$_1$ line (795\,nm laser on the top of Fig.\,\ref{fig:setup}). The two beams are combined by beam splitter BS4, and the quarter-wave plate makes their polarization circular.
The beams are passed through a neutral density filter after which their power is about 2\,mW and expanded by a telescope to a diameter of $\sim$15\,mm. The diameter of the beams that pass through the vapor cell is determined by an iris with a varying diameter.
They pump most of the  $^{87}$Rb atoms to the sub-level $|2,2\rangle$ or to $|2,-2\rangle$, depending on whether the circular polarization is right-handed or left-handed, respectively.

The experimental procedure starts by turning on the static magnetic field and the pumping beams to prepare most of the atomic population at the state $|2\rangle$.
We then use two alternative methods to observe Rabi oscillations \cite{Orr2016}:

Method 1: \textbf{Discrete measurement.}
We turn off the pumping beams and turn on the RF field for a duration $t$ to induce Rabi oscillations.
Then we turn on a linearly polarized probe laser tuned to a frequency of about 200\,MHz above the F=2 $\leftrightarrow$ F'=2 transition in the D$_2$ line (780\,nm laser on the bottom of Fig.\,\ref{fig:setup}). When the polarization of the probe beam is parallel to the magnetic field it  induces only $\pi$ transitions, such that the optical density of the vapor is sensitive to the distribution of the population between the two sub-levels \cite{PhysRevLett.111.053004}.
We repeat this process $n$ times (typically $n$=50), each time increasing  the duration $t$ by $\Delta t$ (typically $\Delta t$=8\,$\mu$s).

Method 2: \textbf{Continuous measurement.}
We install in front of the photodetector (PD) a bandpass filter that transmits  only the power of the 795\,nm  (D$_1$) circularly polarized pumping laser beam passing through the vapor cell. This beam serves as a probe, as the sub-level $|2\rangle$ is a dark state for this beam while $|1\rangle$ absorbs it. We obtain a continuous absorption signal, which monitors the atomic population as a function of time while the RF driving field is on.

Although in the first method only the RF radiation interacts with the atoms during the oscillations while in the second there are two additional light fields, we have verified that the rigidity phenomenon is observed with both methods, as shown by Fig.\,\ref{fig:2methods}(c). In order to quicken data taking thus reducing experimental drifts, we utilize the continuous method throughout this work.

 \subsection{Results: Rabi frequency rigidity}
\label{sec:results}

Figures\,\ref{fig:2methods}(a-b) show the raw data of Rabi oscillations obtained with the above two methods. We extract the Rabi frequency $\omega$ at each RF detuning, as shown in Fig.\,\ref{fig:2methods}(c), by fitting the data to a function
$Ae^{-\gamma t}\cos(\omega t+\phi)+B t+C$,
where $A$ is the oscillation amplitude, $\phi$ is the initial phase, $\gamma$ is the exponential decay rate, $B$ is the drift coefficient of the signal, and $C$ is its baseline. The fitting window is 0.01\,ms$<t<$0.6\,ms, long enough as the oscillations usually vanish before $t$=0.6\,ms.

We compare the measured Rabi frequency derived from the single frequency fit to the oscillation frequency predicted for a homogeneous sample with the same central detuning $\Delta$.
 As noted, $\Omega_R$ should grow with increasing $\Delta$.  However, in contrast and as shown in Fig.\,\ref{fig:2methods}(c), we have unveiled a regime in which the Rabi frequency hardly changes with RF frequency, thereby giving rise to a frequency rigidity.

\begin{figure}
\includegraphics[width=\columnwidth]{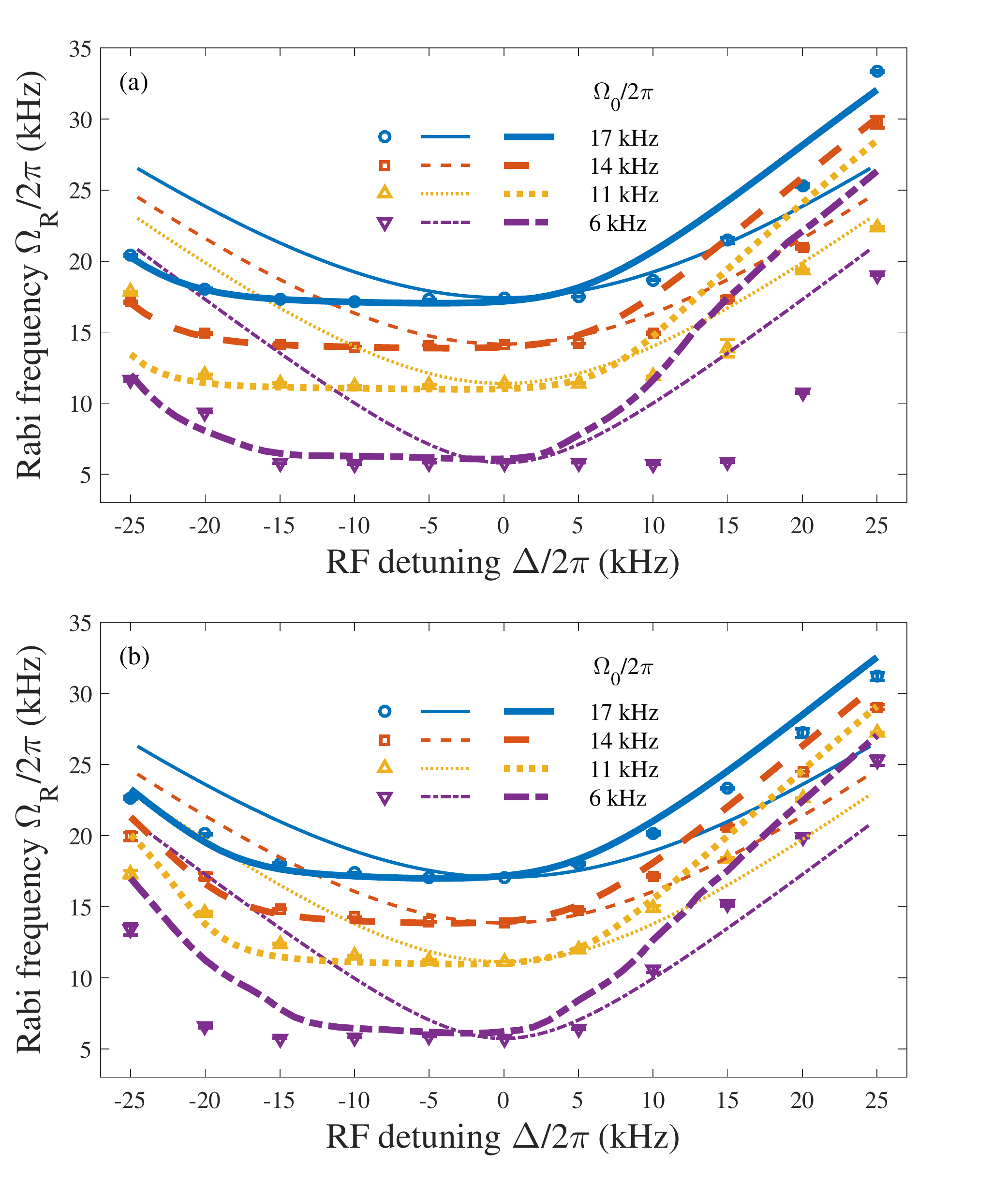}
\caption{(color online) Rigidity of Rabi oscillation frequency for large and small inhomogeneity: experiment (circles, squares, upward and downward triangles) vs. numerical model (thick solid, dashed, dotted and dash-dot lines); thin solid, dashed, dotted and dash-dot lines show the expected $\Omega_{\rm_R}=\sqrt{{\Omega_{0}}^{2}+\Delta^{2}}$. (a) Large magnetic field inhomogeneity (12\,mm probe beam diameter). (b) Small field inhomogeneity (3\,mm diameter). The numerical model assumes a skewed Gaussian distribution of the magnetic field with widths $\sigma/2\pi$=10\,kHz and 8\,kHz for (a) and (b), respectively. In both cases, the skewing is represented by a Gauss error function characterized by erf(-0.3/$\sqrt2$)$\sim$-23.6\%, and the homogeneous decoherence rate is $\gamma/2\pi$=1\,kHz.}
\label{fig:freeze}
\end{figure}

We study this frequency rigidity over a range of RF amplitudes and two sample volumes. As the relative inhomogeneity of the DC magnetic field is small ($\sim$0.05\%) and achieved after a lot of alignment effort, we have chosen not to try to reduce this inhomogeneity by rearranging the magnetic coils system and instead to decrease the inhomogeneity of the sample by decreasing the effective sample volume using a narrower probe beam diameter.
Figs.\,\ref{fig:freeze}(a) and (b) show the observed Rabi frequency for larger and smaller sample volume, respectively, as a function of the RF detuning $\Delta$ from the central resonance frequency of the ensemble. This central resonance is frequency is slowly scanned (see appendix~\ref{sec:asymmetry}, figure~\ref{fig:spectra}).  We collect Rabi oscillation signals for two effective sample volumes by changing the probe/pump beam diameter from 12\,mm to 3\,mm, expecting that the smaller beam diameter pumps and probes atoms over a smaller volume and hence smaller DC magnetic field variation.
Qualitatively, one may observe that the rigidity phenomenon increases with increasing inhomogeneity and decreases with increasing bare Rabi frequency ({\it e.g.} see downward triangle purple data for strongest rigidity).

In what follows we discuss the possible source of the frequency rigidity phenomenon and construct a model that quantitatively explains the results. This model, which was already compared with the experimental results in Figs.~\ref{fig:freeze}, is then compared with additional results concerning more detailed features of the signal, such as its amplitude and frequency spectrum.

\subsection{Interpretation}
\label{sec:numerical_model}

We start by discussing a few possible sources of the frequency rigidity effect other than the inhomogeneity of the magnetic field across the sample and eliminating their relevance to the explanation.

\paragraph{Collisions:} We have performed in the same vapor cell a similar experiment inducing Rabi oscillations using a microwave transition (6.8\,GHz) between non-magnetic levels - the two hyperfine states $|2,0\rangle$ and $|1,0\rangle$ (``clock states").
Rabi oscillations between these levels that are not sensitive to the magnetic field inhomogeneity faithfully satisfied the generalized Rabi frequency rule for bare Rabi frequencies of a few hundreds of Hz~\cite{Amir2007,Gal2009}. This clearly indicates that the frequency rigidity effect is not merely a consequence of intrinsic properties of the atoms in our vapor cell, such as collisions of rubidium atoms with other rubidium atoms or with buffer gas atoms.
In addition, we observed Rabi frequency rigidity for RF transitions in vapor cells with different buffer gases and varying pressures: Ne at 7.5 and 75\,Torr, and Kr at 60\,Torr.
This indicates that collisions are not an essential factor in the explanation, at least not within the range of parameters that was examined.

\paragraph{Doppler broadening:} As the experiment is performed with atoms at room-temperature the Doppler broadening due to thermal velocity distribution in the sample should be considered. However, for a radio-frequency of 18\,MHz, corresponding to the Zeeman splitting due to the 26\,G of the bias magnetic field, the expected Doppler broadening is $\omega_{\rm RF}\cdot\langle v\rangle/c\sim 2\pi\times 15\,$Hz. This is more than two orders of magnitude smaller than the relevant frequency scale, which is determined by the Rabi frequency of the order of 10\,kHz. Furthermore, even 6.8\,GHz microwave transition mentioned above, where the Doppler broadening is on the order of 5\,kHz, much larger than the Rabi frequency, the effect of Doppler broadening on the Rabi oscillations is minor due to the fact that each atom changes its velocity every few nanoseconds by fast collisions with the buffer gas atoms. As the velocity of each atom is randomized at a rate that is much higher than the rate of Rabi oscillations, the process involves Dicke narrowing~\cite{Frueholz1985Dicke}
that eliminates the effect of Doppler broadening such that each atom performs Rabi oscillations as if its velocity is zero.

\paragraph{Damping:} A few damping processes act in our system.
First, thermal relaxation of the population of sub-levels within the ground state of rubidium  due to collisions (a $T_1$ process)
strongly depends on pressure and temperature. $T_1$ was measured in  our system  to be 30\,ms for Ne at 75 Torr,  10\,ms for Ne at 7.5 Torr and 6\,ms for Kr at 60 Torr~\cite{Orr2016}.
These relaxation  rates are not relevant to our results as their time scale is much larger than the 1\,ms duration of Rabi oscillations measurement.
Other damping processes of $T_2$ type due to atomic interactions or damping due to the laser pumping during the oscillations do act within the time scale of the measurement and shown to be on the order of $\tau\sim$0.2\,ms [see Fig.~ \ref{fig:amplitude}(b) below]. However, the damping rates have a strong dependence on the detuning, which indicates that inhomogeneity plays a major role in the process, as we show in section~\ref{sec:theory}. In addition,  damping processes are taken into account within our numerical model presented below. To the best of our knowledge such processes by themselves have never been shown to give rise to a phenomenon similar to the frequency rigidity observed in our experiment.

In order to establish the explanation based on the inhomogeneity of the sample we now set a theoretical model based on the inhomogeneity of the magnetic field across the cell. Note that due to high buffer gas pressure the diffusion length of the atoms during the 1\,ms measurement time is $\sqrt{Dt} \sim 0.5\,$mm \cite{Partniak2014diff} ($D$ is the diffusion constant). Thus, each atom stays in a specific region in the cell with a specific local magnetic field along the time of the measurement.
We may assume that the density of rubidium atoms in the cell is constant, while the magnetic field has a small inhomogeneity $B({\bf r})=B_0+\delta B({\bf r})$, where we are interested only in the absolute value of the field, which determines the Zeeman splitting $\omega_{\rm Zeeman}\approx \frac12\mu_BB({\bf r})/\hbar$. The contribution of different atoms across the sample to the absorption signal is also determined by the profile of the probe beam $I_{\rm probe}({\bf r})$, such that atoms near the center of the probe beam contribute more than those at the edges of the beam.
The absorption signal, proportional to the population in the state $|1\rangle$, is then
\begin{equation}
S\propto \int d^3{\bf r}\, I_{\rm probe}({\bf r})P_1[\Omega_0({\bf r}),\Delta({\bf r}),t], \label{eq:signalr} \end{equation}
where $P_1(\Omega_0,\Delta,t)$ is the population of the Zeeman sub-level $|1\rangle$ as a function of time for a given (local) bare Rabi frequency $\Omega_0$ and detuning $\Delta=\omega_{\rm RF}-\omega_{\rm Zeeman}$, with $\omega_{\rm Zeeman}=\omega_{\rm Zeeman}^{(0)}+\frac12\mu_B\delta B({\bf r})/\hbar$ depends on the local magnetic field. For a pure two-level system prepared at the state $|2\rangle$ we have $P_1(\Omega_0,\Delta,t)=(\Omega_0/\Omega_R)^2\sin^2(\Omega_R t/2)$, where $\Omega_R=\sqrt{\Omega_0^2+\Delta^2}$ is the generalized Rabi frequency. In the case of our 5-level $F=2$ manifold this form may be also affected by transitions into other Zeeman sub-levels and by the pumping effect of the probe laser beam or other $T_1$ ad $T_2$ processes.

We have independently measured the inhomogeneity of the magnetic field across the cell and found that $\delta B\sim 15\,$mG, corresponding to a Zeeman shift of $\delta\omega_{\rm Zeeman}\sim 2\pi\times 10\,$kHz. Although the relative inhomogeneity is quite small, $\delta B/B_0\sim 0.05$\% the spectral inhomogeneous width $\frac12\mu_B\delta B/2\pi\hbar\sim 10$\,kHz is on the same order as the bare Rabi frequency $\Omega_0$. In contrast, we have found the inhomogeneity of the bare Rabi frequency itself by measuring the variation of the amplitude of the RF field to be $\delta\Omega_0/\Omega_0\sim$2\%, so that the variation of $\Omega_0$ is much smaller than the 10\,kHz order of the relevant frequencies involved in the Rabi oscillation.
Therefore we may neglect the effect of this inhomogeneity and  consider $\Omega_0$ to be constant over the sample.

Our independent measurement of the magnetic field inhomogeneity provides a rough estimation of its magnitude but does not provide a detailed distribution of the field. In order to provide a model for the inhomogeneity we use the Dirac $\delta$ function: we multiply both sides of Eq.~(\ref{eq:signalr}) by $\int db\, \delta[b-B({\bf r})]$ and integrate over the possible values $b$ of the static magnetic field.
We then obtain
\begin{equation}
S\propto \int db\, \rho(b)P_1(\Omega_0,\Delta (b),t), \label{eq:Bdist}
\end{equation}
where  $\rho(b)=\int d^3{\bf r}\, I_{\rm probe}({\bf r})\delta[b-B({\bf r})]$ is the distribution of the static magnetic field over the probe beam profile. The expression in Eq.~(\ref{eq:Bdist}) is now independent of the unknown explicit form of the spatial distribution of the magnetic field and we may model the distribution $\rho(b)$ of the contribution of atoms in a specific field $b$ to the signal by giving it simple forms.

We calculate the occupation $P_1(\Omega_0,\Delta,t)$ for a given RF field amplitude and detuning by solving the master equations for the five Zeeman sub-levels of the F=2 hyperfine state with the driving field and a damping parameter $\gamma$ representing the $T_1$ and $T_2$ times in the presence of the probe/pump beam.
The occupation $P_1$ is then convolved with a model of the distribution $\rho(b)$, which we take to be a skewed Gaussian parameterized by the spectral width $\sigma=(\mu_B/2\hbar)\delta b$ of the magnetic field inhomogeneity over the sample. The skewness of the Gaussian is introduced in order to account for asymmetry of the Rabi rigidity effect for blue-detuned RF fields and red-detuned RF field, as observed in Fig.\,\ref{fig:freeze}).
We then fit curves of the population vs. time resulting from the integral in Eq.~(\ref{eq:Bdist}) in the same way that we fit the experimental data to extract the Rabi oscillation frequency $\omega$ for various values of the central detuning, RF field amplitudes and spectral widths of the inhomogeneity (determined by the probe beam diameter.

Our model is found to be in good agreement with the experimental results, as shown in Figs.~\ref{fig:freeze}(a-b).
An inhomogeneity of 10\,kHz, (corresponding to the independently measured inhomogeneity of 15\,mG out of 26\,G  fits the experimental results for the 12\,mm beam (Fig.~\ref{fig:freeze}).
For the 3\,mm beam,  the best-fit inhomogeneity is only reduced to 8\,kHz  (less than the 4-fold decrease from 12 to 3\,mm) since the inhomogeneity is mostly along the axial direction of the cell.
Reducing the beam diameters further is impractical, due to a decrease of the observable signal.
One may also consider reducing the rigidity by increasing the driving field amplitude instead, which however leads in our specific case to a breakdown of the two-level approximation by introducing a two-photon transition to a third level.

Beyond the success of the numerical model to replicate the experimental results, we also wish to gain understanding of the mechanism that leads to the rigidity phenomenon.
We suggest that whenever the inhomogeneous distribution includes values of the static magnetic field corresponding to a Zeeman splitting resonant with the driving field ({\it i.e.} when $\sigma\gtrsim \Delta$) , the Rabi oscillations of this resonant part of the ensemble are dominant in the overall oscillation signal. This dominance of the part oscillating with a frequency $\Omega_0$ is not only due to the amplitude of these oscillations, which  is larger than those far from resonance (by a Lorentzian factor $1/(1+\Delta^2/\Omega_0^2)$, but also because off-resonant oscillations of different parts of the ensemble interfere destructively and give rise to a signal that is damped very quickly. We will demonstrate and explain this process more clearly with our theoretical insight presented in section~\ref{sec:theory}.

We can now explain the apparent asymmetry of frequency rigidity In Figs.~\ref{fig:freeze}(a-b).
The distribution $\rho(b)$ is wider towards weaker static magnetic fields (lower transition frequencies), such that when the driving field frequency is red-shifted from the central resonance ($\Delta<0$) there are still many atoms in a region where the magnetic field is low and corresponds to a transition frequency near resonance with the driving field. In contrast, when the driving field is blue-detuned with respect to the overall transition frequency only a small number of atoms are near resonance with the driving field as the static field distribution is narrower towards stronger fields, and consequently the surviving signal at the bare Rabi frequency is not observed.
This asymmetry is related to the specific configuration of our system and is not essential to the frequency rigidity phenomenon, as discussed in appendix~\ref{sec:asymmetry}.

\subsection{Additional properties of the phenomenon}

\begin{figure}
\centering
\includegraphics[width=\columnwidth]{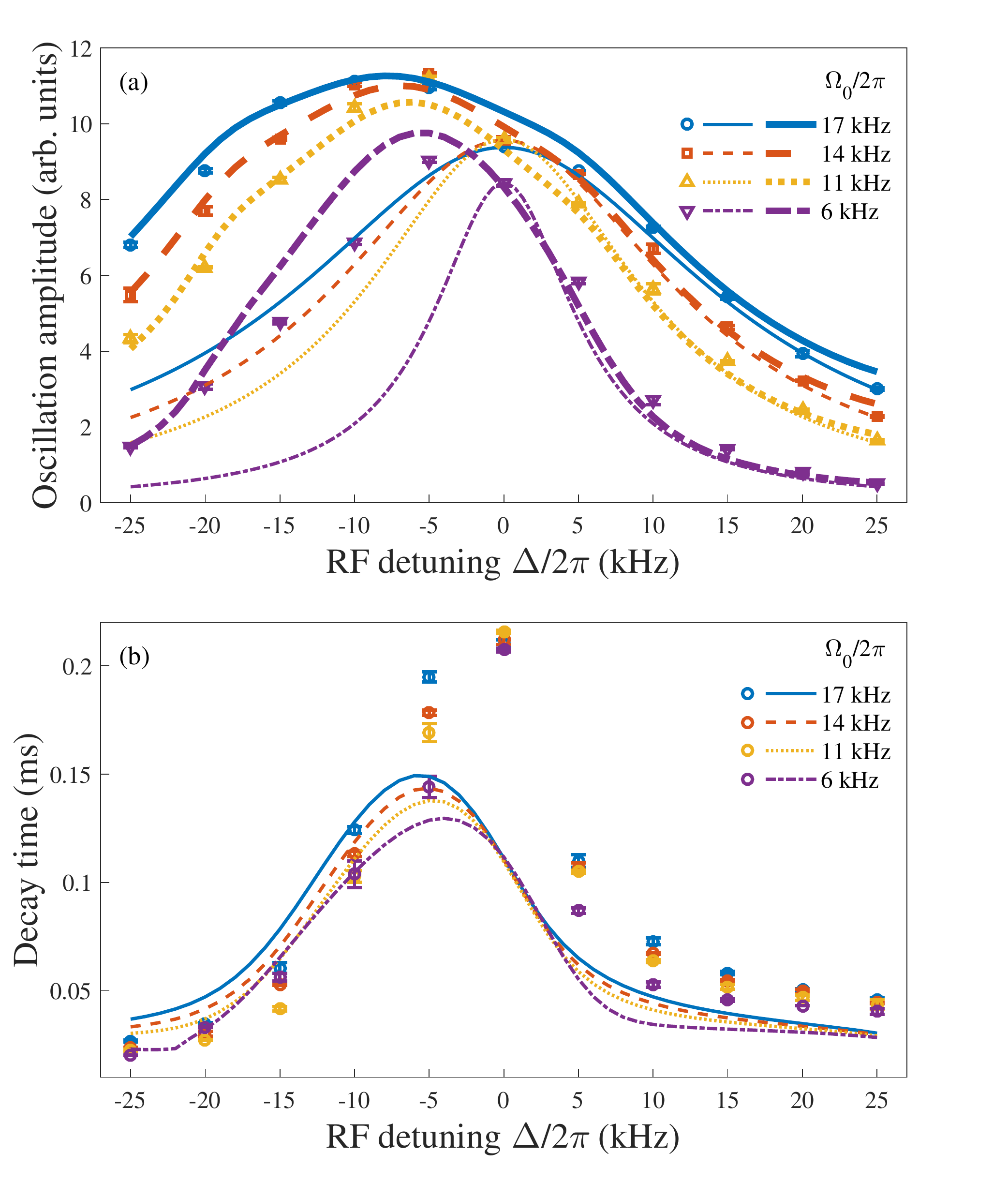}
\caption{(color online) Rabi oscillation amplitude vs. RF detuning for different RF field intensities (in terms of driving rate $\Omega_0$).
(a) Shows the initial amplitude $A$ and (b) shows the decay time $\tau=1/\gamma$ of the oscillation amplitude.
Circles, squares, upward and doward triangles are the values extracted from the experimental data used in Fig.~\ref{fig:freeze}(b), thin (solid, dashed, dotted and dash-dot) lines in (a) are based on a homogeneous model $A\propto 1/(1+\Delta^2/\Omega_0^2)$  and thick (solid, dashed, dotted and dash-dot) lines are based on our numerical model with the same parameters as in Fig.~\ref{fig:freeze} (small inhomogeneity). }
\label{fig:amplitude}
\end{figure}

The magnetic field inhomogeneity does not only affect the frequency of Rabi oscillation signal but also its amplitude: both the initial amplitude of the oscillations  (parameter $A$ in the fit presented in the beginning of section~\ref{sec:results}) and the decay time of the oscillation amplitude ($\tau=1/\gamma$).
In Fig.~\ref{fig:amplitude}(a) we present the initial amplitude of oscillations  as a function of detuning $\Delta$ for  different RF field intensities. In a homogeneous sample we expect the amplitude to be a Lorentzian function of the detuning: $A(\Delta)\propto 1/(1+\Delta^2/\Omega_0^2)$ (thin solid, dashed, dotted and dash-dot lines). However, in the presence of inhomogeneity the amplitude is a weighted average as in Eq.~(\ref{eq:Bdist}). In Fig.~\ref{fig:amplitude}(a) we compare
the experimental results (circles, squares, upward and downward triangles) to the numerical model based on such an average and find a good agreement. The asymmetry of the magnetic field distribution is responsible to an apparent broadening of the curve of the amplitude towards red detuning as observed above for the frequency rigidity.
In Fig.~\ref{fig:amplitude}(b) we present the decay time of the oscillations as a function of  detuning. The decay time is determined both by the intrinsic decay due to $T_1$ and $T_2$ processes, including the effect of the probe/pump beam and the effect of inhomogeneity. As we show in section~\ref{sec:theory} the inhomogeneity of the magnetic field induces damping due to destructive interference between different oscillation frequencies in the ensemble, which increases for larger detunings $|\Delta|$. Good agreement between experimental data and simulation is shown in both figures~\ref{fig:amplitude}(a) and (b).  The slight disagreement in the decay time in (b) may follow from the incompleteness of the theoretical model and would be cured by including more details in the model, such as independent values for $T_1$ and $T_2$ or a more detailed shape of the magnetic field distribution, which we wanted to avoid in order to emphasize the generality of the rigidity phenomenon.

\begin{figure}
\includegraphics[width=\columnwidth]{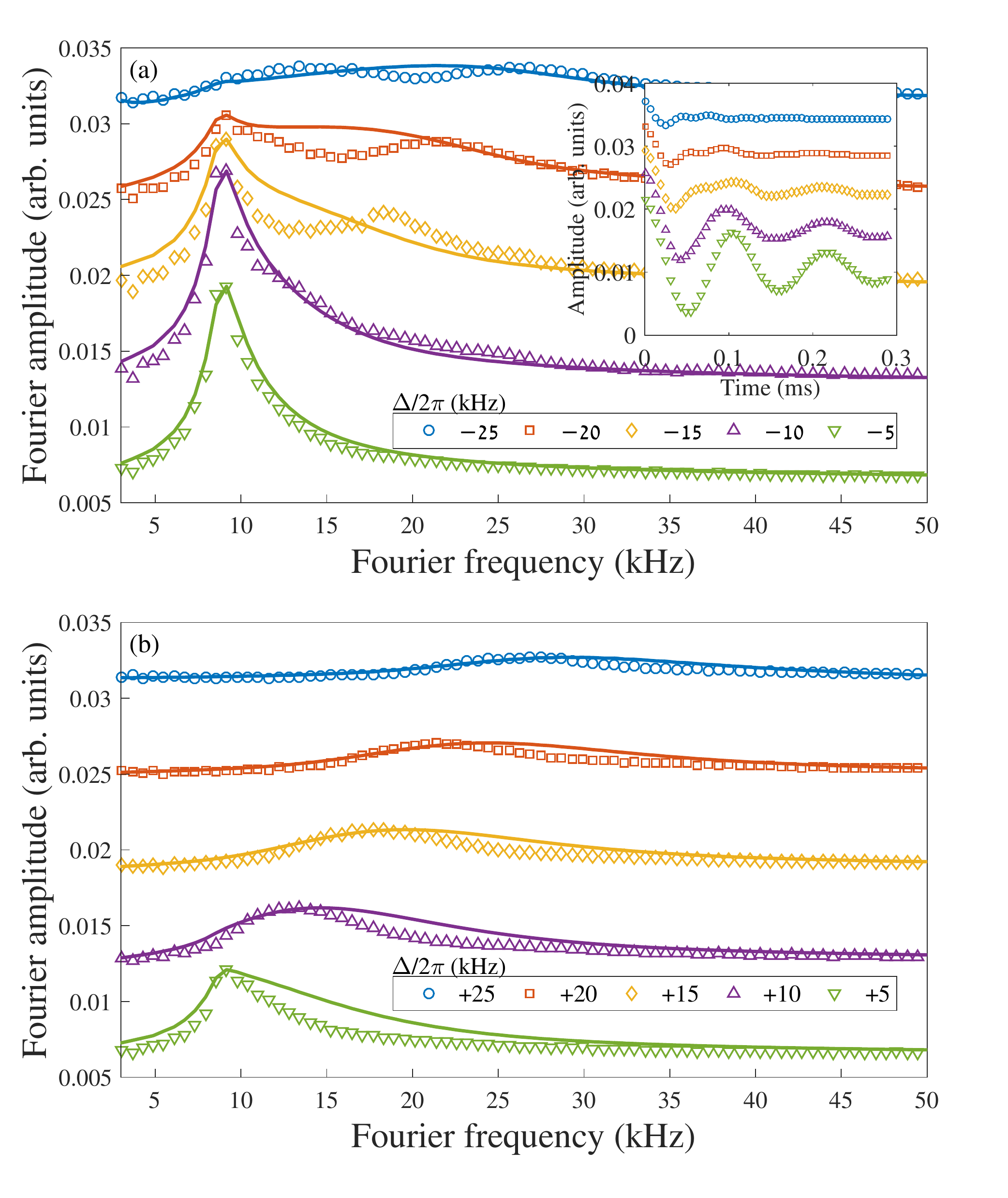}
\caption{(color online) Fourier analysis of the observed Rabi oscillations: experiment (circles, squares, bottom and up triangles) vs. numerical model (solid lines) for a red (a) and blue (b) detuned RF field. In (a) the Fourier spectra shows a double peak structure. As explained in the text, the two peaks correspond to two oscillation modes: high frequency - fast decaying, and low frequency - slow decaying. These cause the time-domain frequency shift shown in the inset of (a). The rigidity is presented in (a) by a sharp peak at $\Omega_{0}/2\pi\sim$9\,kHz, unchanged regardless of the increasing $\Delta$. Parameters used in the numerical model are the same as for Fig.\,\ref{fig:freeze}(b). Each Fourier curve is displaced vertically for clarity of presentation.}
\label{fig:FFTspectra}
\end{figure}

A phenomenon directly related to the frequency rigidity is a shift from higher to lower oscillation frequency during the oscillation. This phenomenon appears more in our measurements for certain experimental parameters, as demonstrated in the inset of Fig.\,\ref{fig:FFTspectra}(a), for red detuning and a 3\,mm probe beam. We observe a frequency shift, where at first the Rabi oscillations have a high frequency and then after a few cycles a lower frequency dominates. We analyze the data by a Fourier transform, which is shown in Fig.\,\ref{fig:FFTspectra}(a) (circles, squares, upward and downward triangles). The two-peak structure demonstrates the frequency shift. The rigidity is due to the dominance of the low frequency component, which maintains the same frequency (bare Rabi frequency) independent of the detuning. In contrast, in Fig.\,\ref{fig:FFTspectra}(b), the Fourier spectra of blue-detuned Rabi oscillations show that the rigidity disappears, as in Fig.\,\ref{fig:freeze}. Solid lines in Fig.\,\ref{fig:FFTspectra} (main panels) show the results of the numerical model for the same parameters. These results
qualitatively agrees with the experiment. The two-peak structure in the numerical results is not as clear as in the experimental data.
We believe that the discrepancy is due to a complicated magnetic field distribution, which could not be simulated by our simple model.
The numerical results indicate that in general the frequency shift occurs for a wide range of parameters and may be either fast or slow. However, due to our limited signal-to-noise ratio we were able to observe it clearly in the experiment except for the case of small inhomogeneity (3\,mm beam) and red-detuned RF, as shown in the inset of Fig.~\ref{fig:FFTspectra}(a). In section~\ref{sec:theory} we present a simple toy model that demonstrates that for a wide range of parameters the signal can be represented by a fast decaying component at a high frequency and then a slowly damped signal at the bare Rabi frequency, hence the frequency shift is inherent to the rigidity phenomenon.

\section{Theoretical insight}
\label{sec:theory}

In this section we present a simplified model of a two-level system driven by an external field and show how Rabi frequency rigidity emerges from the inhomogeneity of the resonance frequency across the sample. The model presented here attempts to capture the fundamental features of the phenomenon in order to gain more understanding of its mechanism and provide a basis for further exploration of the possible appearance of the phenomenon in other systems rather than the one studied in the experiment.

\subsection{Basic model}

Consider a sample of many two-level atoms with an inhomogeneous energy splitting between their levels $|1\rangle$ and $|2\rangle$.
The atoms are initially in level $|2\rangle$ and Rabi oscillations are induced by a driving field with a well-defined frequency detuned from the atomic transition frequency by $\delta$. We assume that the value of $\delta$ varies across the sample due to the inhomogeneity of the transition frequency and has a distribution with an average $\langle \delta\rangle=\Delta$. The average detuning $\Delta$ can be controlled by varying the frequency of the driving field. For simplicity we assume a symmetric Gaussian distribution of $\delta$ with a spectral width $\sigma$.
The Gaussian distribution serves as a convenient model for the inhomogeneity.

It is important to emphasize that the main features of the resulting effect are not restricted to this specific shape, as they follow from more general principles explained below.
In addition, we note that the source of inhomogeneity of the detuning frequency $\delta$ may be of different kinds: either a local variation of an electric or magnetic field that shifts the atomic levels or a Doppler shift due to thermal motion. However, here we consider the case where the frequency shifts are static and do not change during the oscillation. This implies that our model does not account for a Doppler shift due to an atomic velocity that changes during the oscillation due to collisions or for fast moving atoms that change their local field during their motion. In the specific case of our experiment, which used an atomic vapor with a buffer gas, the atoms stay at the same location during the measurement and the Doppler broadening is negligible, as discussed in section~\ref{sec:numerical_model},
so that the model presented here is relevant to the experiment and simplifies the more detailed model presented above.

Our model has only three parameters: $\Delta$, $\sigma$ and the bare Rabi frequency $\Omega_0$ determined by the driving field amplitude and the coupling of the field to an atom, which are assumed not to vary across the sample.
Note that this simplified model does not include intrinsic damping processes ($T_1$ and $T_2$) as in our numerical model of section~\ref{sec:numerical_model} so that decay of the resulting Rabi oscillation signal will be solely due to the inhomogeneity.
The integrated Rabi oscillation signal is an expression equivalent to Eq.~(\ref{eq:Bdist}), specifically for this model
\begin{equation}
S(t)=\frac{1}{2\sqrt{2\pi}\sigma}\int d\delta\, e^{-(\delta-\Delta)^2/2\sigma^2} \frac{1-\cos[\omega(\delta)t] }{1+\delta^2/{\Omega_0}^2},
\label{eq:densities}
\end{equation}
 where  $\omega(\delta)=\sqrt{\Omega_0^2+\delta^2}$ is the generalized
Rabi frequency for a atoms with a given $\delta$. As noted, we model the distribution of transition frequencies by a simple symmetric Gaussian function.

The integral in Eq.~(\ref{eq:densities}) is not in general solvable analytically. In order to gain insight into the behavior of the signal
let us first consider the case of a small inhomogeneity $\sigma\ll \Omega_0$.
In this case the Lorentzian factor $(1+\delta^2/\Omega_0^2)^{-1}\approx (1+\Delta^2/\Omega_0^2)^{-1}$ can be taken out of the integral and the generalized Rabi frequency can be approximated by $\omega(\delta)\approx \Omega_R+[\Delta/\Omega_R(\Delta)](\delta-\Delta)$, where $\Omega_R=\omega(\Delta)$. The
integral is then given by
\begin{eqnarray}
&&\frac{1}{\sqrt{2\pi}\sigma}\frac{1}{1+\Delta^2/\Omega_0^2}\int d\xi\, e^{-\xi^2/2\sigma^2}\times
\label{eq:subsignal} \\
&&\times\left\{1-\cos\left[\left(\Omega_R+\frac{\Delta}{\Omega_R}\xi\right)t\right]\right\} \nonumber \\
&&= \frac{1}{1+\Delta^2/\Omega_0^2}\left[1-\cos(\Omega_Rt)\exp\left(-\frac{\sigma^2\Delta^2}{2\Omega_R^2}t^2\right)\right],
\nonumber \end{eqnarray}
representing an oscillation that decays with a rate $\gamma(\Delta)=\sigma\Delta/\Omega_R$ ranging from $\gamma=0$ for $\Delta=0$ to $\gamma=\sigma$ for $\Delta\gg \Omega_0$.

Let us now consider a sample with an arbitrarily large inhomogeneity $\sigma$. We may look at the oscillation signal in this sample as consisting of the sum of partial signals from sub-samples with small spectral widths $\bar{\sigma}$, each giving rise to a signal of the form of Eq.~(\ref{eq:subsignal}) for different detuning frequencies, $S(\Delta,t)\propto\sum_j A(\Delta_j)\{1-\cos[\omega(\Delta_j)t] e^{-\gamma_J^2 t^2/2}\}$, where $A(\Delta_j)= e^{-(\Delta_j-\Delta)^2/2\sigma^2} /(1+\Delta_j^2/\Omega_0^2)$ is a product of the (Gaussian) abundance of atoms with detuning $\Delta_j$ in the sample and the (Lorentzian) oscillation amplitude of these atoms.
At short times $t<\gamma_j^{-1}$ the dominant oscillations are determined by $A(\Delta_j)$, giving some more weight to oscillations closer to resonance, but also to those at the central detuning.
However, after a long time the contribution of sub-samples with large detuning $|\Delta_j|$  decays and the only oscillation frequencies that survive are those due to detunings near $\Delta_j\sim 0$, whose decay rate is small, $\gamma_j\sim 0$. This gives rise to the frequency rigidity effect and the frequency shift from high oscillation frequency in the beginning of the oscillation to the bare Rabi frequency at long times. The exponential decay of the sub-ensembles with large detuning frequencies is a consequence of the inhomogeneity and represents a mechanism involving destructive interference between frequency components with high variation at long times.

Note that the above explanation of the effect is only qualitative, as the division of the sample into subgroups is arbitrary. The spectral width of each sub-group and hence the decay rate of its partial signal is not uniquely determined and the actual mechanism of the formation of the signal involves interference between all the contributions from different atoms. A more rigorous explanation based on a numerical fitting procedure is presented below.

\subsection{Two-frequency fit }
\label{sec:2freq}

By performing the integral in Eq.~(\ref{eq:densities}) numerically we find that for almost the whole parameter space the oscillation signal can be very well fitted (up to a constant) to a two-frequency model
\begin{equation}
S(t)\approx Ae^{-\gamma_a^2t^2/2}\cos(\Omega_{0} t+\phi_a)+Be^{-     \gamma_b^2t^2/2}\cos(\bar{\Omega}t+\phi_b),
\label{eq:fit} \end{equation}
representing two oscillation modes emerging from different parts of the sample: one with a low frequency $\Omega_0$ and a slow decay rate $\gamma_a$, and one with a higher frequency $\bar{\Omega}$ and a fast decay rate $\gamma_b$ ($\gamma_b\gg \gamma_a$).
The Gaussian decay is in line with the Gaussian form of the distribution but other decay forms are expected if the distribution has a different form (for example, exponential decay for a Lorentzian distribution).
It follows that
the Rabi oscillation signal becomes after some time $t\sim \gamma_b^{-1}$ dominated by oscillations at approximately the bare Rabi frequency $\Omega_0$ even if the expected ``typical" oscillation frequency of the central part of the sample is the generalized Rabi frequency $\Omega_R$, which may be much larger than $\Omega_0$ for $\Delta>\Omega_0$.

\begin{figure}
\includegraphics[width=\columnwidth,trim={1cm, 1cm, 1cm, 1cm}]{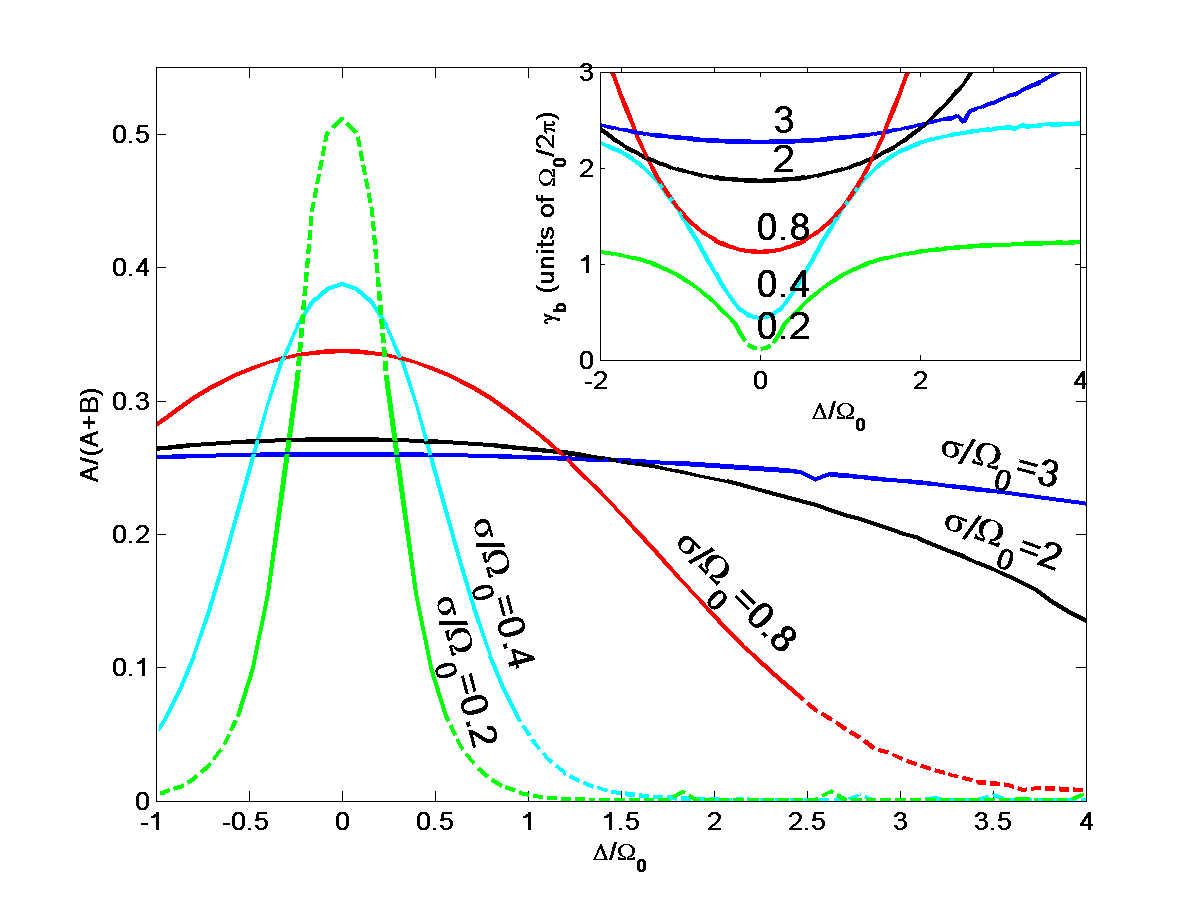}
\caption{(color online) A two-frequency fit for the abstract model of rigidity: initial fraction of the signal oscillating with the bare Rabi frequency $\Omega_0$ vs. driving field detuning (in units of $\Omega_0$) for varying inhomogeneity $\sigma$ of the transition frequency  (also in units of $\Omega_0$). For $\sigma>\Omega_0$ the origin of the rigidity phenomenon becomes evident. Inset: Decay rate $\gamma_b$ of the high frequency part of the Rabi oscillation as a function of detuning at the same inhomogeneity values (taking $\gamma_a=0$).
Solid (dashed) parts of the curves indicate that the full width of the 95\% confidence interval for the fit parameters is smaller (larger) than 25\% of their values. }
\label{fig:fracA}
\end{figure}

To study the interplay between inhomogeneity and driving rate, we calculate the parameters of Rabi oscillations following from the fit in Eq.~(\ref{eq:fit}) for different ratios $\sigma/\Omega_0$
As shown in Fig.~\ref{fig:fracA}.
The fit was performed for times $0<t<10\times 2\pi/\Omega_0$ and we have fixed $\gamma_a$ to be zero. The goodness of fit (using the $r^2$ test) ranges between 99\% for $\sigma/\Omega_0=0.2$ to 94\% for $\sigma/\Omega_0=3$ in the range of $\Delta$ shown in the figure.
When $\sigma\ll \Omega_0$ there is no rigidity as $A\rightarrow 0$; the fraction oscillating with $\Omega_0$ is high only when the driving field frequency is near resonance, where $\Omega_{\rm_R}\approx\Omega_0$.
In this case the two-frequency fit becomes redundant and the fit parameters become highly uncertain (denoted by a dashed curve for uncertainty larger than 50\%).
However, when $\sigma>\Omega_{0}$ a large fraction of the ensemble oscillates with $\Omega_{0}$ at any detuning
and its oscillation, having a slow decay constant, completely dominates at longer times.
In the inset we show the decay rate of the high frequency mode. For $\sigma\geq \Omega_0$ the decay rate $\gamma_b>\Omega_0$ corresponds to the damping of this mode in a time $\tau\sim \gamma_b^{-1}\ll 2\pi/\Omega_0$, namely in a time shorter than one oscillation period,
so that for almost the whole time the signal is dominated by the frequency $\Omega_0$ and the shift from high to low oscillation frequency is not easily visible.
Note that the simple model used in this section does not include damping due to other processes rather than the inhomogeneity itself, so that a quantitative comparison of the results shown in the figure with the experimental results are not instructive.

\section{Outlook and conclusions}
\label{sec:summary}

Here we have described a general phenomenon that is likely to appear in many kinds of inhomogeneous systems performing driven oscillations. If the dependence of the oscillation frequency $\Omega$ on the inhomogeneous factor (let us call it $\eta$) that varies across the sample has an extremum at a given value of $\eta$, where $\partial\Omega/\partial\eta=0$, then an effect of frequency rigidity similar to what we observed in this work may appear. The overall (average) oscillation signal  in the long time range would then become dominated by a single frequency that does not change if the inhomogeneous factor $\eta$ is shifted over the whole sample. We have demonstrated this frequency rigidity for Rabi oscillations where the detuning parameter, which is controlled by the frequency of the external driving field, is varying across the sample. The rigidity effect follows from the fact that the dependence of the Rabi frequency on the detuning $\delta$ has a minimum at $\delta=0$, as explained in section~\ref{sec:theory}, giving rise to rigidity at the bare Rabi frequency.

We have shown theoretically and experimentally how the frequency rigidity effect emerges from the interplay between driving rate and inhomogeneous broadening and analyzed in detail some more specific features of this phenomenon.
 Our analysis provides a quantitative guide concerning the interplay between inhomogeneity and driving rate (driving field power).
The effects are universal and do not depend on the specific type of two-level system, the origin of inhomogeneity of the resonance frequency, or the specific form of interaction with the driving field.
In devices for which miniature dimensions are required, as well as fast operations, keeping clear of the restrictions described here may not be easy to achieve.

Some surprising protocols may arise; for example, in the case of an inhomogeneity that cannot be suppressed to a low enough level, it is advantageous to increase the inhomogeneity so that $\gamma_b$ becomes larger and the Rabi oscillation becomes monochromatic (bare frequency) faster (although the signal becomes weaker in analogy to ``Doppler-free" spectroscopy). This gives, for example, an accurate measure of the driving field intensity at the position of the ensemble.

\begin{acknowledgements}

We thank Mark Keil for his critical reading of the manuscript. This work is funded in part by the Israeli Science Foundation, the EC Matter--Wave consortium (FP7--ICT--601180), and the German-Israeli DIP project (Hybrid devices: FO 703/2--1) supported by the DFG.
\end{acknowledgements}

\appendix

\section{Asymmetry effect }
\label{sec:asymmetry}

In this section we briefly present and explain the effect of asymmetry observed in Figs.~\ref{fig:freeze} and~\ref{fig:amplitude}. We believe that this is not a fundamental effect since, for example, it is hardly observable in the $|2,-2\rangle$ $\leftrightarrow$ $|2,-1\rangle$ transition, as described in the following. Instead, as shown, it stands to reason that the effect follows from a specific configuration of the experimental setup responsible for a different magnetic field distribution across the cell in the two measurement configurations.
We present these much ``cleaner" results only here, towards the end of this article, as this was the order of observation in the lab, and as we wanted to demonstrate that our numerical model can also account for experimental imperfections.

\begin{figure}
\centering
\includegraphics[width=\columnwidth]{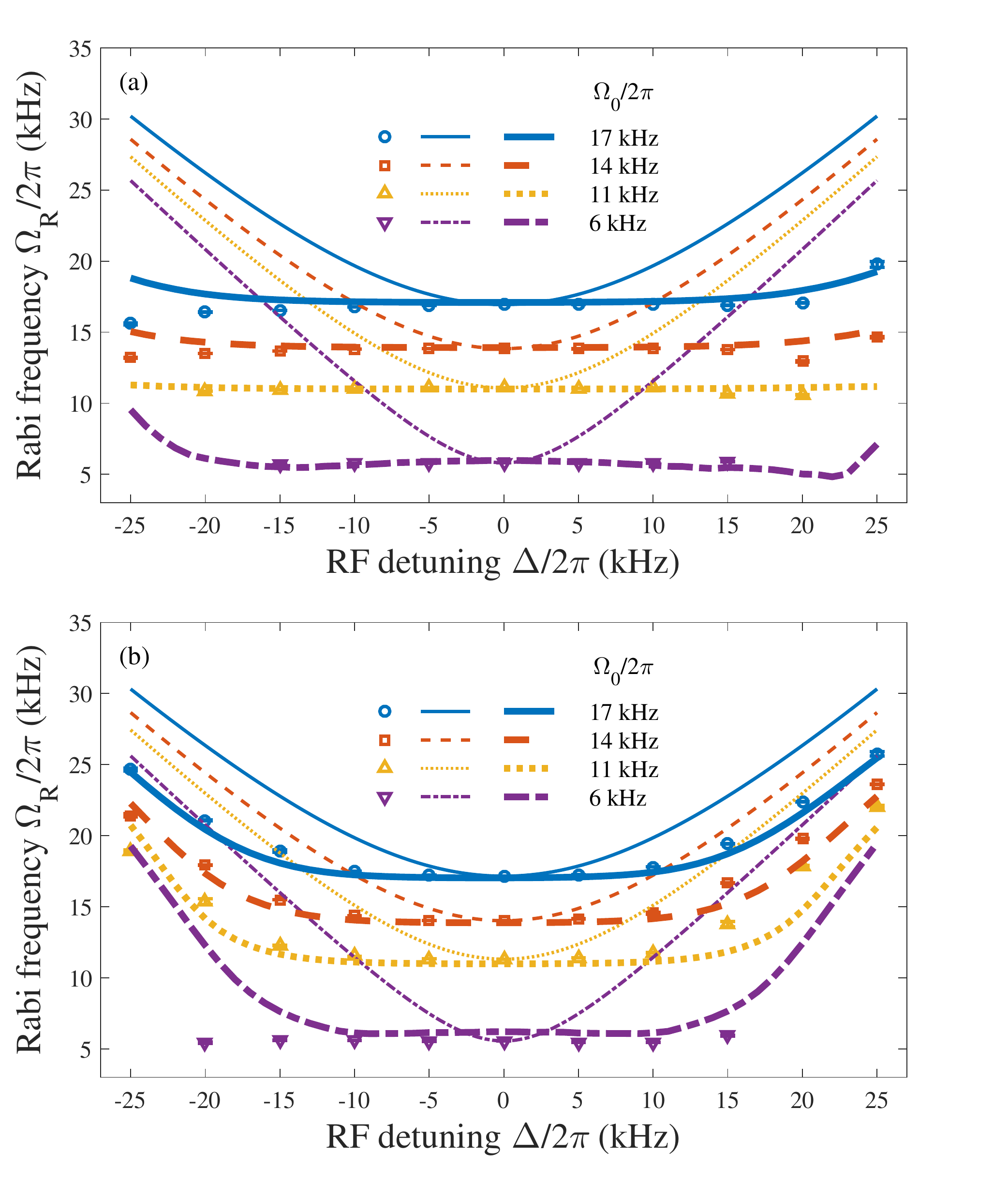}
\caption{(color online) Rigidity of the Rabi oscillation frequency for large and small inhomogeneity in the $|2,-2\rangle$ $\leftrightarrow$ $|2,-1\rangle$ transition. Presented are the experimental results (circles, squares, upward and downward triangles), the expected $\Omega_{\rm R}=\sqrt{\Omega_0^2+\Delta^2}$ (thin solid, dashed, dotted and dash-dot lines) and the numerical model (thick solid, dashed, dotted and dash-dot lines).  (a) Large magnetic field inhomogeneity (beam diameters of 12\,mm). (b) Small field inhomogeneity (beam diameters of 3\,mm). The numerical model assumes a symmetric Gaussian distribution of the magnetic field with widths $\sigma/2\pi$=12\,kHz and 8\,kHz for (a) and (b), respectively, and a homogeneous decoherence rate of $\gamma/2\pi$=1\,kHz.}.
\label{fig:minus2transition}
\end{figure}

Figure~\ref{fig:freeze} shows that the frequency of Rabi oscillations between the sub-levels $|2,2\rangle$ and $|2,1\rangle$ behaves in a different way for red and blue detuning. For comparison, we present in Fig.~\ref{fig:minus2transition}
 a similar plot for the transition $|2,-2\rangle \leftrightarrow |2,-1\rangle$, in which the asymmetry between red and blue detuning is very slight.

\begin{figure}
\includegraphics[width=\columnwidth]{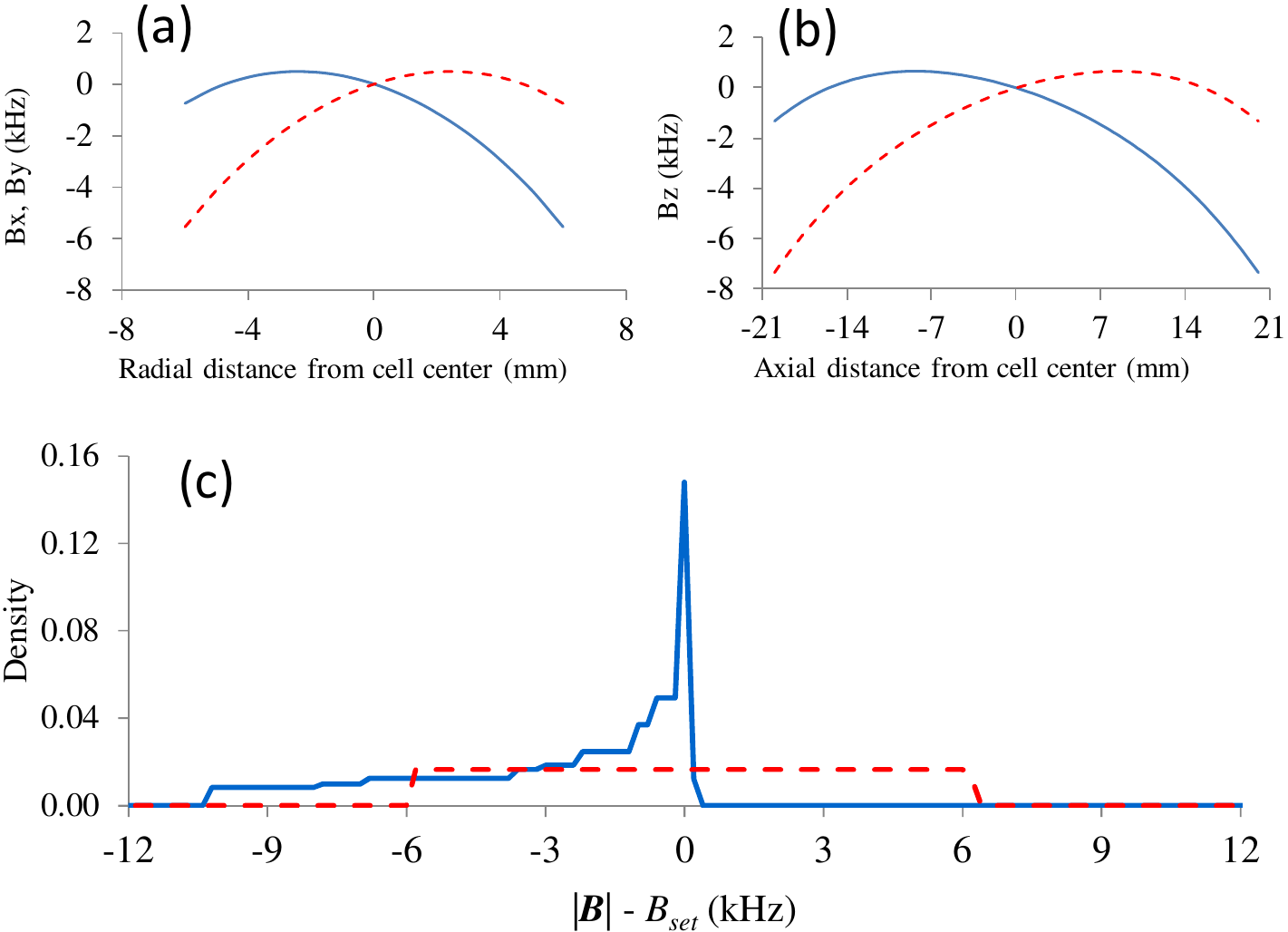}
\centering
\caption{(color online) (a-b) An example for the possible dependence of the components of the DC magnetic field (in kHz, 10\,mG=7\,kHz) inside the cell as a function of position. (a) Radial components: solid blue line - $B_{0x,y}$; dashed red line - $B_{1x,y}$ (see text). (b) Axial components: solid blue line - $B_{0z}$; dashed red line - $B_{1z}-B_{set}$.
(c) Distribution of the magnetic field deviation (in kHz, 10\,mG=7\,kHz). Dashed red - $|\mathbf B_{tot+}|-B_{set}$. (Average 0.1 kHz, 49.4\% below average, 50.6\% above average); Solid blue - $|\mathbf B_{tot-}|-B_{set}$. (Average -3.16 kHz,  42\% below average, 58\% above average).}
\label{fig:field_dist}
\end{figure}

A possible explanation for the difference between the asymmetry in the two transitions lies in the magnetic configuration used for each of them. Switching between these configurations is done by reversing the current in the $z$-axis bias coils (see Fig.~\ref{fig:setup}).
As demonstrated below, reversing the current can change the magnetic field distribution around its target value $B_{set}$. The symmetry of the magnetic field distribution, which was modeled by skewness of the Gaussian distribution in our numerical model used for fitting the experimental results, determines also the symmetry or asymmetry of the Rabi frequency dependence on the detuning, as demonstrated in Fig.~\ref{fig:minus2transition} (symmetric) and Fig.~\ref{fig:freeze} (asymmetric).

The details of the deviation of the magnetic field from its target value are unknown to us. However, we can use a simple model to demonstrate how the symmetry of the distribution can change when the magnetic field due to the current in the $z$ bias coils is reversed. In general, the magnetic field is a sum of two parts: the field $\mathbf {B_0}\sim (0,0,0)$ due to earth and the compensation coils and $\mathbf {B_1}\sim(0,0,\pm B_{set})$ due to the $z$ bias coils. Our model assumes for simplicity that each component $B_x$,$B_y$ and $B_z$ of the two parts of the field varies only along the respective directions $x$, $y$ and $z$ ($-20\,\rm{mm}<z<20\,\rm{mm}$ along the axis of the cell and $-8\,\rm{mm}<x,y<8\,\rm{mm}$ along the transverse direction).

The field components for this example are presented in Fig.~\ref{fig:field_dist}(a-b). The magnetic field is given in units of kHz (700\,kHz/G). The variation of the magnetic field  is well within $\pm$8\,kHz  ($\pm0.05\%$ of $B_{set}$).
The distributions of the values of the total magnetic field $|\mathbf B_{tot\pm}|=|\mathbf B_0 \pm  \mathbf B_1|$ calculated at a grid of points (spacing: 0.5\,mm) in the active volume of the vapor cell are presented in Fig.~\ref{fig:field_dist}(c). Clearly, reversing the current can change a symmetric distribution to a skewed one.

\begin{figure}
\centering
\includegraphics[width=\columnwidth]{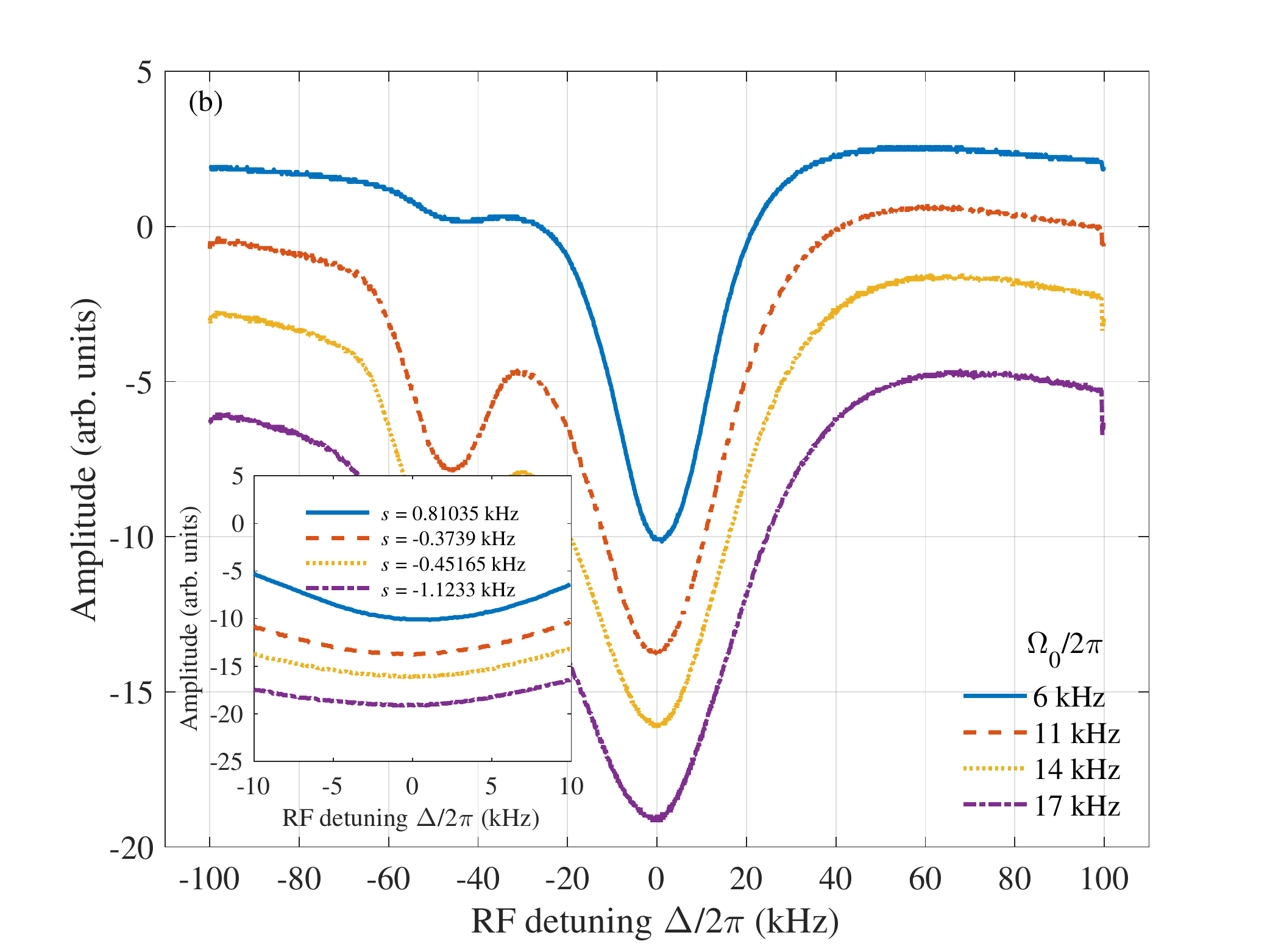}\\
\includegraphics[width=\columnwidth]{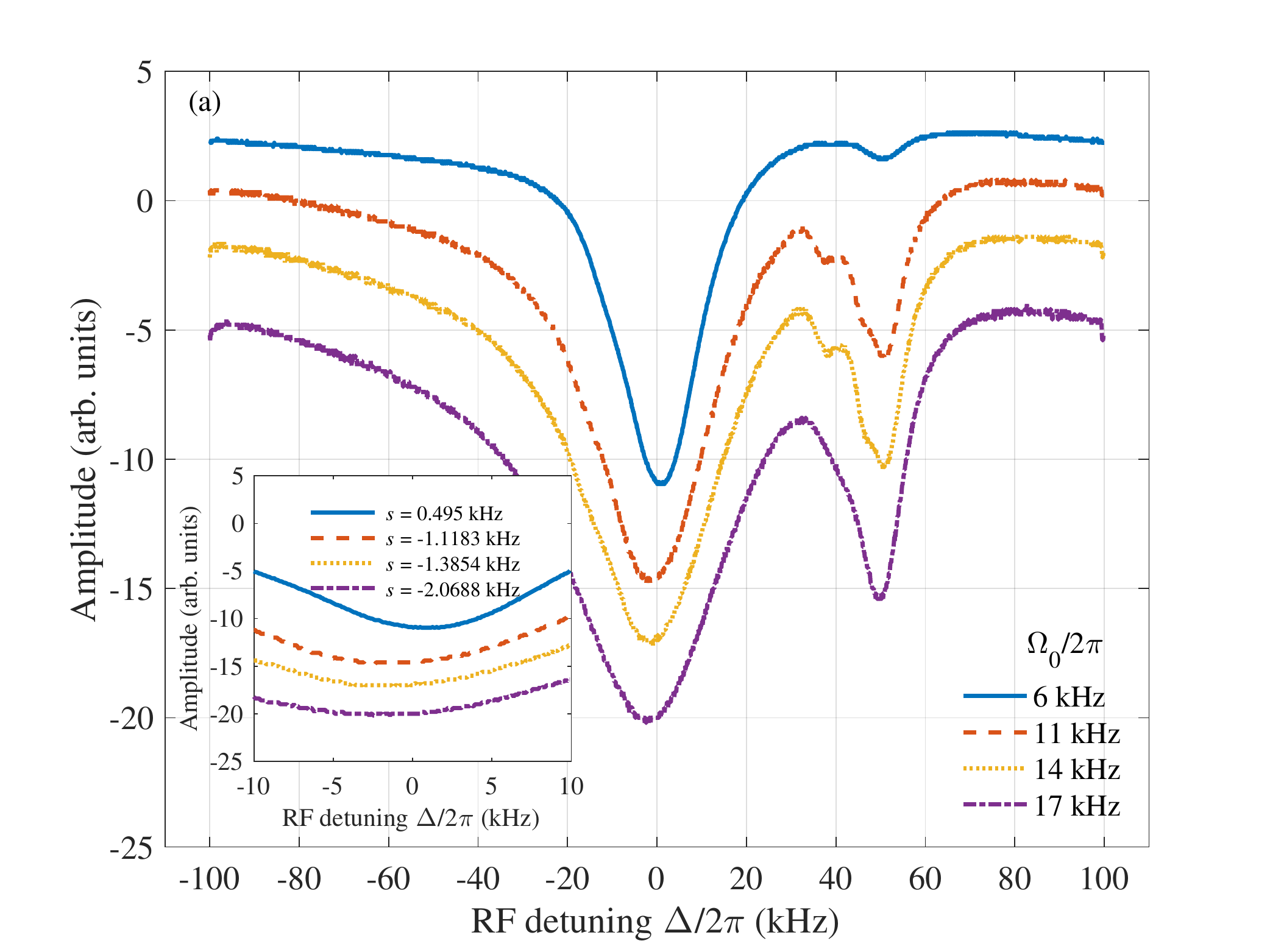}
\caption{(color online) Spectra of the laser power exiting the vapor cell as a function of the detuning $\Delta$.  The RF power level for each spectrum is given by the Rabi frequency  it induces ($\Omega_0$).  The main dip in (a) represents the $|2,2\rangle$ $\leftrightarrow$ $|2,1\rangle$ transition (18.167\,MHz) and in (b) the transition $|2,-2\rangle$ $\leftrightarrow$ $|2,-1\rangle$ (18.481\,MHz). The smaller dip at a detuning of $\pm$50\, kHz in (a) and (b) is the two-photon $|2,\pm 2\rangle$ $\leftrightarrow$ $|2,0\rangle$ transition. The spectra are shifted  from each other in the vertical direction for clarity. In the inset we present a a zoom-in on the dip area, and list the deviations $s$ of the dips associated with different RF power.}
\label{fig:spectra}
\end{figure}

Finally, the asymmetry effect may be related to another observation made in the process of measuring the atomic population while slowly scanning the RF frequency in the presence of the DC magnetic field and the pumping beams (method 2 described in section~\ref{sec:setup}). The RF frequency was scanned during 1\,s in a window of $\pm 100\,$kHz across the resonance transition frequency as initially calculated from the estimated DC magnetic field in the vapor cell.
Figure~\ref{fig:spectra}(a) presents such spectra for the $|2,2\rangle$ $\leftrightarrow$ $|2,1\rangle$ transition at various values of the RF power, and Fig.~\ref{fig:spectra}(b) presents similar spectra for the $|2,-2\rangle$ $\leftrightarrow$ $|2,-1\rangle$ transition.  The spectra represent the pump beam transmission for the steady-state optical density in the presence of both a pump and a driving RF field. We define the resonant transition frequency to be used as the reference frequency throughout this work as the RF frequency at the minimum of the spectrum measured for a RF power corresponding to a Rabi frequency of $\Omega_0/2\pi= 9\,$kHz. In our setup, we find the resonance frequency to be 18.167 MHz for the $|2,2\rangle$ $\leftrightarrow$ $|2,1\rangle$ transition and 18.481 MHz for the $|2,-2\rangle$ $\leftrightarrow$ $|2,-1\rangle$ transition.
As seen by zooming-in on the dips, the position of the minimum changes with RF power. We have fitted a Voigt profile \cite{Rotonardo1997} to the spectra to locate the exact value of the minima and find that they are detuned from resonance by -2 to 0.5\,kHz for the $|2,2\rangle$ $\leftrightarrow|2,1\rangle$ transition and by  -1.1 to 0.8\,kHz for the $|2,-2\rangle$ $\leftrightarrow|2,-1\rangle$ transition.

 These slight changes in the resonance frequency and the detunings of the minima again indicate that the magnetic field in both configurations is different. The detailed modelling and understanding of these slight changes is beyond the scope of this paper, as it requires exact knowledge of the magnetic field which is not available to us.

\end{document}